\documentclass[%
reprint,
groupedaddress,
amsmath,amssymb, 
aip,
article,pre,
]{revtex4-2}
\usepackage{epsfig,amsmath,amsfonts,amssymb}
\usepackage{graphicx}
\usepackage[utf8]{inputenc}
\usepackage{epsfig,psfrag}
\usepackage{mathbbol}
\usepackage{multirow}
\usepackage{tikz}
\usepackage{hyperref}
\usepackage{bm}

\newcommand{\beq}{\begin{equation}}
\newcommand{\eeq}{\end{equation}}

\newcommand{\beqa}{\begin{eqnarray}}
\newcommand{\eeqa}{\end{eqnarray}}

\newcommand{\opt}{\text{opt}}

\begin{document}
\title{Active charge and discharge of a capacitor: scaling solution and energy optimization}
\author{S. Faure}
\affiliation{Laboratoire de Collisions Agr\'egats R\'eactivit\'e,
CNRS UMR 5589, FeRMI, Universit\'e Paul Sabatier, 118 Route de
Narbonne, 31062 Toulouse CEDEX 4, France}
\author{C. A. Plata}
\affiliation{Física Teórica, Universidad de Sevilla, Apartado de
 Correos 1065, E-41080 Sevilla, Spain}
\author{A. Prados}
\affiliation{Física Teórica, Universidad de Sevilla, Apartado de
 Correos 1065, E-41080 Sevilla, Spain}
\author{ D. Gu\'ery-Odelin}
\affiliation{Laboratoire de Collisions Agr\'egats R\'eactivit\'e,
CNRS UMR 5589, FeRMI, Universit\'e Paul Sabatier, 118 Route de
Narbonne, 31062 Toulouse CEDEX 4, France}

 \date{\today}
\begin{abstract}
  Capacitors are ubiquitous in electronic and electrical devices. In
  this article, we study---both theoretically and
    experimentally---the charging and discharging of capacitors using
  active control of a voltage source. The energy of these processes is
  analyzed in terms of work and heat. We show how to approach the
  quasistatic regime by slowing down the charging or discharging
  processes. Conversely, we study the price to be paid in terms of
  Joule heat when we speed up these processes. Finally, we
  develop optimal processes that minimize energy consumption
  for a finite charging time. Our work
    combines fundamental concepts from thermodynamics, classical
    mechanics and electrical circuits, thus blurring the artificial
    frontiers at the undergraduate level between these
    disciplines. Also, it provides a simple example of a
    prominent problem in current science, the optimization of
    energy resources. Moreover, our study lends itself well to an
  experimental project in the classroom, involving computer
  control of a voltage source, data acquisition, and processing.
\end{abstract}

\maketitle

\section{Introduction}\label{sec:intro}

In our teaching of elementary physics, we divide physics into
  separate boxes to address different aspects of our description of
  physical systems. Two of these typical boxes are classical
  mechanics, to study the dynamical evolution of mechanical systems,
  and thermodynamics, to study the equilibrium states of macroscopic
  systems. As a consequence of our presentation, students often see
  these boxes as completely separate, with little or no connection
  between them.

On the one hand, at the undergraduate level, thermodynamics
  is basically a static theory. Yet, we introduce the idea of
  thermodynamic processes, in which some externally controlled
  parameter---e.g. the volume of the container---is continuously
  changed, but in such a way that the system is always at
  equilibrium. Therefore, time plays no role in these
  \textit{quasistatic processes}, although we typically tell students
  that a quasistatic process is an idealization, an infinitely slow
  process in which an infinitesimal change of
  the externally controlled parameter involves an infinite
  time.\cite{callen_thermodynamics_1985}

On the other hand, in undergraduate courses of physics,
  variational calculus is typically associated with our building of
  Lagrangian mechanics. The minimization of the action, which is the
  time integral of the system Lagrangian, leads to the well-known
  Lagrange equations, which are second-order ordinary differential
  equations for the time evolution of the generalized coordinates in
  configuration
  space.\cite{goldstein_mechanics_1980,landau_mechanics_1976} Often,
  we stress the generality of the mathematical framework of this
  minimization problem to students---stating that it is not restricted
  to classical mechanics.  However, this generality statement is
  somehow compromised by
  our applying this framework exclusively to mechanical problems. 

Currently, the thermodynamics of irreversible processes of
  mesoscopic systems is a hot topic of research. In this field,
  thermodynamic quantities can be introduced for systems with few
  degrees of freedom: this is the realm of stochastic
  thermodynamics,\cite{sekimoto_stochastic_2010} in which fluctuations
  play a key role. At this level of description, the introduction of
  heat and work are quite transparent and, in agreement with the ideas
  introduced in the undergraduate thermodynamics courses, they are
  functionals of the irreversible process followed by the system. For
  given initial and final states and a connection time, we may try to
  devise a suitable protocol---a swift state-to-state
  transformation---for the driving that makes the desired
  connection,\cite{guery-odelin_driving_2023} i.e. a control problem
  arises. Then, there appears the problem of evaluating the cost, in
  terms of dissipation, of these irreversible connections. Quite
  naturally, there also appears the problem of finding the optimal
  protocol that minimizes the
  dissipation,\cite{guery-odelin_driving_2023,blaber_optimal_2023}
  i.e. an optimal control problem arises. On the mathematical side,
  this optimal control problem is akin to the variational calculus
  problem in classical mechanics.\cite{liberzon_calculus_2012} On the
  physical side, there are interesting consequences of the
  minimization procedure, such as the emergence of speed
  limits---i.e. the existence of a minimum operation time for a given
  value of the
  dissipation.\cite{schmiedl_optimal_2007,aurell_optimal_2011,sivak_thermodynamic_2012}
  Moreover, the idea of quasistatic process appears also naturally:
  for obtaining zero dissipation, the operation time must diverge.

The ideas presented in the above paragraph are appealing from
  a pedagogical point of view, since they allow us to introduce dynamical
  ideas in a thermodynamic problem, basic notions of control theory,
  and present the power of variational calculus in a context that is
  different from the usual one. In addition, they are useful to
  clarify how irreversibility is inevitably linked to dissipation, and
  the trade-off between operation time and dissipation expressed by
  the speed limits. Nevertheless, the complexity of the mathematical
  framework of stochastic thermodynamics framework makes it impossible
  to use these ideas directly in the undergraduate classroom.

In this work, we consider a simple physical system, an RC
  circuit, to illustrate the above ideas in a minimal setup and in a
  pedagogical way. Using
  such a basic circuit, which can be certainly employed at the
  undergraduate level, allows us to introduce irreversible processes
  and identify thermodynamic quantities. This is possible despite the
  system not being explicitly coupled to a heat bath---which, on the
  other hand, avoids the problem of stochasticity. We note that the RC
  circuit under study constitutes a minimal device for energy storage,
  which makes it relevant for introducing the stimulating topic of the
  optimal use of energy resources in the classroom.  

We address the problem of active charge and discharge of a capacitor, that is, we are interested in devising an \textit{active} input source to drive the system to a target charge state instead of predicting the \textit{passive} response to a given input.
  The simplicity of the system dynamics enables the
  mathematical treatment of the control problem. 
  Importing simple
  ideas from the so-called fast-forward
  protocols,\cite{masuda_fast-forward_2010,guery-odelin_shortcuts_2019,
    nakamura_fast_2020,plata_taming_2021,impens_time_2021,guery-odelin_driving_2023}
  we engineer finite-time connections and analyze their thermodynamic
  balance. Also, we address the problem of minimizing
  the dissipation, which leads to a simple Euler-Lagrange
  equation. Physical consequences of this equation, like the trade-off
  between time operation and dissipation encoded in the speed limits,
  are also discussed.

\section{Model}\label{sec:model}

\begin{figure}
    \begin{center}
        \includegraphics[width=6cm]{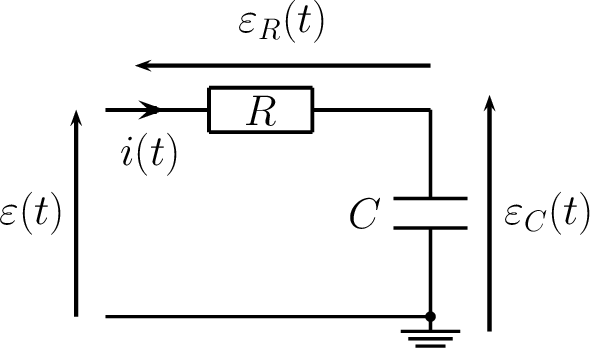}
    \end{center}
    \caption{RC circuit under a properly engineered voltage source.
      The total time-dependent voltage is $\varepsilon(t)$, the current intensity is $i(t)$, and the voltage drop in the resistance and the capacitor are $\varepsilon_{R}(t)$ and $\varepsilon_{C}(t)$, respectively.
    }
    \label{fig1}
\end{figure}
We consider the electric  circuit depicted in Fig.~\ref{fig1}, consisting of a resistor placed in series with a capacitor driven by a time-dependent voltage $\varepsilon(t)$. The sum of voltages around a loop is zero, as a result
 \begin{equation}
 \label{eq:kir}
    \varepsilon(t)=\varepsilon_{R}(t)+\varepsilon_{C}(t), 
  \end{equation}
where $\varepsilon_{R}(t)\equiv R \,i(t)$ and $\varepsilon_{C}(t)=q(t)/C $ are the voltage drops in the resistor and the capacitor, respectively, and $i(t)\equiv \dot{q}(t)$ is the current intensity. The charge in the capacitor $q(t)$ obeys the differential equation:
\begin{equation}
\dot q(t) + \frac{q(t)}{\tau} = \frac{\varepsilon(t)}{R},
\label{eqrc}
\end{equation}
with the characteristic time
\begin{equation}\label{eq:tau-def}
  \tau=RC.
\end{equation}

The charging of a capacitor under a constant voltage
$\varepsilon(t)=\varepsilon_f$  is readily obtained by solving
Eq.~(\ref{eqrc}) with the initial condition $q(0)=0$. The charge then
evolves according to {the time-dependent function}
\begin{equation}\label{eq:q(t)-ref-proc}
  q_r(t) = C\varepsilon_f (1-e^{-t/\tau}).
\end{equation}
We have introduced the subindex $r$ in $q_r(t)$ to mark that it will be
used as the reference {process} in what follows. From a fundamental point
of view, this reference charging process of the capacitor lasts an
infinite time, the charge increases monotonically but only reaches its
final value
  \begin{equation}\label{eq:qf-def}
    q_f \equiv C \varepsilon_f
  \end{equation}
  asymptotically for $t\to\infty$,
  $q_{r}(\infty)\equiv\lim_{t\to\infty}q_{r}(t)=q_{f}$. From a
  practical point of view, the capacitor can be considered as fully
  charged for $t\gtrsim 5\tau$, for which $q_{r}(t)\gtrsim 0.99
  q_{f}$.


\section{Scaling solutions}\label{sec:scaling-solutions}

Inspired by fast-forward techniques recently developed in a wide variety of domains,\cite{masuda_fast-forward_2010,guery-odelin_shortcuts_2019, nakamura_fast_2020,plata_taming_2021,impens_time_2021,guery-odelin_driving_2023} we consider the charging function $q(t) = q_r(\Lambda(t))$, which is essentially the time rescaling of the reference process $q_r(t)$. The rescaling function $\Lambda(t)$ effectively allows the system to go through the same intermediate states as the original process but at a different rate.  For sake of simplicity, we use the linear time rescaling function $\Lambda(t) = \alpha t$. Depending on the value of $\alpha$, the charging process can be accelerated ($\alpha>1$), slowed down ($0<\alpha<1$), frozen ($\alpha=0$), or even reversed ($\alpha<0$).
\begin{figure}
\centering
\includegraphics[width=8.cm]{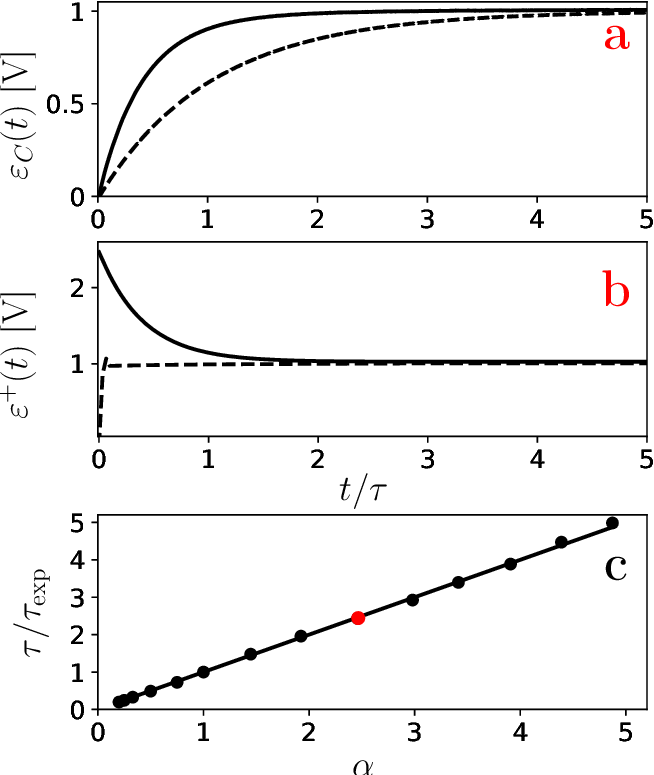}
\caption{Charge of a capacitor. (a) Measured voltage
  $\varepsilon_C(t)$ for a charge under a constant voltage
  $\varepsilon_f=1\,$V (dashed line), and for the scaled protocol with
  speed-up factor $\alpha=2.5$ (solid line). (b) Measured driving
  voltage $\varepsilon(t)$, with the same code as in (a). (c) Ratio
  $\tau/\tau_{\rm exp}$ as a function of the scaling factor, where
  $\tau_{\rm exp}$ is obtained by an  exponential fit of the measured
  voltage $\varepsilon_C(t)$ depicted in (a). The  $\alpha=2.5$ data
  is represented as a red disk, and is associated to the data
  represented with a solid line in (a) and (b). We observe the expected
  acceleration of the charging process. Initial condition:
  $\varepsilon_C(0)=0$ V.}
\label{FigureScaling1}
\end{figure}

By substituting the rescaled solution $q(t)$,
  \begin{equation}
    \label{eq:q-rescaled}
    q(t)=C \varepsilon_{f} \left(1-e^{-\alpha t/\tau}\right)
  \end{equation}
into Eq.~(\ref{eqrc}), we deduce the required voltage driving to generate such a solution:
\begin{equation}
\varepsilon^{+}(t) = R \left(\frac{dq}{dt} + \frac{q}{\tau} \right) = \varepsilon_f + \varepsilon_f(\alpha-1)e^{-\alpha t/\tau}.
\label{eq3}
\end{equation}
We use the symbol $+$ (resp. $-$) for the charging (resp. discharging) process.
As expected, the voltage to be applied remains unchanged for $\alpha=1$. Note that we use here an inverse engineering approach in which the driving is inferred by imposing the ``trajectory''.\cite{guery-odelin_shortcuts_2019,faure_shortcut_2019,guery-odelin_driving_2023}

The corresponding experiment has been conducted with
$R=1296$~$\Omega$, $C=47$~$\mu$F ($\tau=60.9$~ms), and
$\varepsilon_f=1$~V. For all experimental data associated with the
charging process, $\varepsilon_{C}(0)=0$~V since the capacitor is
initially discharged. To drive the voltage source, $\varepsilon(t)$,
with the appropriate time dependency, we use the LabVIEW control of a waveform generator Keysight 33611A 80 MHz, which allows the generation of arbitrary time dependent input voltage. In practice, we discretize the theoretical curve in a succession of voltage steps with a time step of $\Delta t=100$~$\mu$s.
 The response is measured over the capacitor voltage
$\varepsilon_{C}(t)$, which is recorded using a Rhode et Schwartz RTM3004
oscilloscope. Since the dynamics is deterministic (noise is
negligible), the results are obtained by recording a single
trajectory. A typical data collection consists of a time series of about 50000 recorded values with the same time steps as for the voltage synthesis {$\Delta t\;\simeq  1.6\times 10^{-3}\tau$}. Precision and accuracy has been tested allowing to neglect error bars. For the charging process, results are summarized in Fig.~\ref{FigureScaling1}.

\begin{figure}
\centering
\includegraphics[width=8.cm]{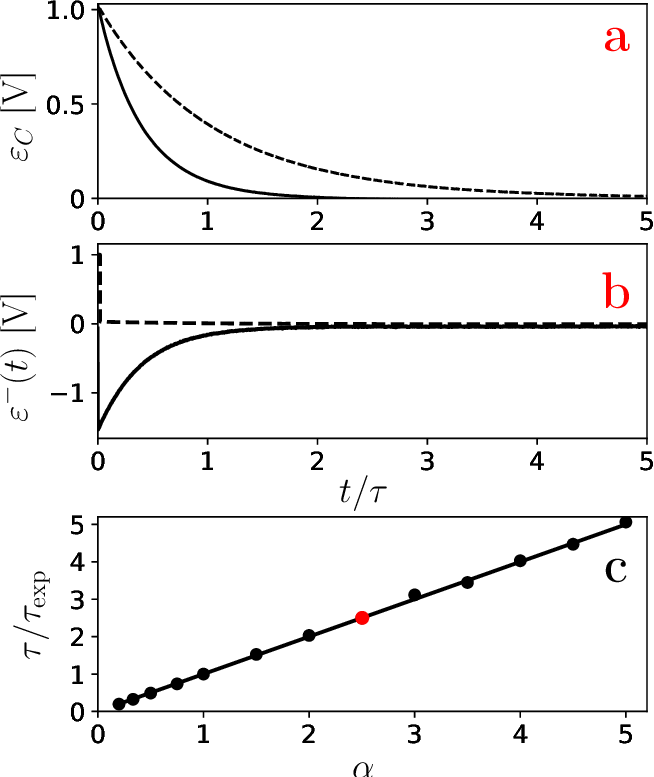}
\caption{\emph{Discharge of a capacitor}. Same notations as Fig.~\ref{FigureScaling1}. The initial condition for the
    discharging process is $\varepsilon_{C}(0)=1$~V, i.e., the final
    situation for the previous charging process in
    Fig.~\ref{FigureScaling1}.
  }
\label{FigureScaling2}
\end{figure}

We measure the voltage $\varepsilon_{C}(t)$ to monitor the dynamics of the capacitor process. In Fig.~\ref{FigureScaling1}a, we present two examples of such measurements. We plot $\varepsilon_{C}(t)$ for the reference charging process ($\alpha=1$, dashed line) and for a rescaled case with $\alpha=2.5$ (solid line). The imposed voltage is represented in Fig.~\ref{FigureScaling1}b.
Additionally, we measure the voltage $\varepsilon(t)$ over time to
verify its proximity to the desired engineered voltage. It is
consistently very close to the one programmed in the voltage
synthesizer, except for very short times---see
Figs. \ref{FigureScaling1}b and \ref{FigureScaling2}b. The slight initial discrepancy in the reference step protocol from the expected driving is due to the rapid variation of the voltage. This indicates the presence of a finite voltage slew rate, which has a more significant impact on the result when the $\alpha$ parameter is large. Nonetheless, we observe a good agreement between our experimental results and the theory, with discrepancies typically below 5\%.

The voltage $\varepsilon_{C}(t)$ is subsequently fitted with an exponential function of characteristic time $\tau_{\exp}$, $\varepsilon_{C}(t) = \varepsilon_f (1-e^{-t/\tau_{\exp}})$. Equation~\eqref{eq:q-rescaled} entails that the theoretical prediction is $\varepsilon_{C}(t)=\varepsilon_f \left(1-e^{-\alpha t/\tau}\right)$, i.e., the theoretical characteristic time is $\tau/\alpha$. We clearly observe the acceleration of the reference process as expected. We have repeated this experiment and analysis for 13 values of the scaling factor $\alpha$ ranging from 0.2 to 5. Figure~\ref{FigureScaling1}c confirms the acceleration ($\alpha >1$) or deceleration ($\alpha <1$) of the reference process as expected through the linear relation $\tau/\tau_{\exp}=\alpha$ (solid line).

Similar analysis and calculations can be easily adapted for the discharge of a capacitor having an initial charge $q(0)$. Specifically, we find:
\begin{equation}\label{eq:eps-}
\varepsilon^{-}(t) = -\frac{q(0)}{C} \left( \alpha -1\right)e^{-\alpha t/\tau}.
\end{equation}
The corresponding experimental results are summarized in
Fig.~\ref{FigureScaling2} for $q(0)=C\varepsilon_{C}(0)$, with
$\varepsilon_{C}(0)=1$~V. {Note that the initial state for the
discharging process is the final state for the previous charging protocol.}


\section{Energy balance}\label{sec:energy-balance}

The energy delivered by the generator over a given time interval $[0,t_f]$ is 
\begin{equation}
\label{eq:defW}
  W = \int_0^{t_f} dt\, \varepsilon(t)\,i(t).
\end{equation}
This expression can be usefully rewritten by using the differential equation~\eqref{eqrc}:
\begin{eqnarray}
W  & = & \int_0^{t_f}  \left[ \frac{q}{C} + Ri\right] i(t)dt \,\nonumber \\
   & = & \frac{q^2(t_f)-q^2(0)}{2C} + R \int_0^{t_f} i^2(t) dt \nonumber \\
  & = & \Delta U - Q, \label{eq:first-principle}
\end{eqnarray}
where we have defined
\begin{align}
\label{eq:ener-heat}
  &\Delta U\equiv \frac{q^2(t_f)-q^2(0)}{2C}, & Q\equiv-  R \int_0^{t_f} dt \, i^2(t) 
\end{align}
In this form, the two terms in Eq.~\eqref{eq:first-principle} have a
clear physical interpretation. On the one hand, the first term is
independent of the ``trajectory'' followed to charge the capacitor: it
is the change of energy stored in the capacitor, which only depends on
the initial and final conditions, thus our notation $\Delta U$. On the
other hand, although our evolution equations are deterministic and do
not incorporate the coupling to a heat bath, we can identify the
second term with the heat $-Q$ that is released to the environment. Of
course, this second term depends on the trajectory and equals the
energy dissipated in the resistor, which is always
positive.\footnote{Note that we have written the first law as
  $\Delta U = Q+W$, i.e. we follow the usual sign convention in which
  energy exchanges, in the form of either heat or work, increasing
  (decreasing) the internal energy of the system are positive
  (negative). Hence, dissipation involves a negative $Q$, which
  motivates our writing of the non-negative dissipated heat as
  $-Q$.}

Note that Eq.~\eqref{eq:first-principle} is the first law of
thermodynamics, $\Delta U=Q+W$; sometimes also called the first
principle. Therefore, $W$ is the work done on the system by the
external source, which is the time integral of the power
$\varepsilon(t)i(t)$, which makes our notation consistent. In the
mechanical analog of the RC circuit, $q(t)\to x(t)$, with $x$ being
the particle's position, and $\varepsilon(t)\to F(t)$, with $F$ being
the external force applied to the particle, so
$W\to \int_{0}^{t_{f}} F(t) \dot{x}(t) dt$ is, in effect, the work
done on the particle.

Let us  apply this thermodynamic perspective to the reference process. The energy input by the voltage generator when it delivers a constant voltage $\varepsilon_f$ to reach asymptotically the charge $q(\infty)=C\varepsilon_f$ is:
\begin{equation}
  W_{r}^{+}=\varepsilon_f \int_0^{\infty} \frac{dq_r}{dt} dt = C \varepsilon_f^2.
\end{equation}
In the charging process, the change of internal energy is
\begin{equation}
    \Delta U^{+}=\frac{q(\infty)^{2}}{2C}=\frac{C \varepsilon_{f}^{2}}{2}.
\end{equation}
Therefore, one has that
\begin{equation}
    -Q_{r}^{+}=W_{r}^{+}-\Delta U=\frac{C \varepsilon_{f}^{2}}{2}.
\end{equation}
 In other words, when a capacitor is charged under a constant voltage, half of the energy input by the generator is dissipated in the resistor.

If we want to accelerate (or decelerate) the charge or discharge by a factor $\alpha$ using the previous scaling solutions, the energy delivered by the voltage generator is readily calculated:
\begin{equation}
  W^{\pm}(\alpha) = \Delta U^{\pm}-Q^{\pm}(\alpha)=\frac{C\varepsilon_f^2}{2} (\pm 1+\alpha).
\label{energypm}
\end{equation}
The first term is the change of internal energy in the charging/discharging process, $\Delta U^{\pm}=\pm C \varepsilon_{f}^{2}/2$, which is logically independent of the acceleration factor $\alpha$. Note that, for the sake of keeping a simple notation in the discharging process, $\varepsilon_f$ is the initial voltage but not the final, which is zero. A complete general framework, including partial charging and discharging processes, is provided in Appendix~\ref{app:partial-charg-discharg}. The second term is the heat released to the environment,
\begin{equation}\label{eq:heat-func-alpha}
    Q^{\pm}(\alpha)=-\alpha \frac{C\varepsilon_f^2}{2} =\alpha Q_{r}^{\pm}.
\end{equation}
For the reference discharging process, $\alpha$=1, the energy
  supplied by the voltage generator vanishes since the voltage is set
  at 0 V for $t>0$, i.e., $\varepsilon^{-}(t)=0$, as given by
  Eq.~\eqref{eq:eps-}. All the energy stored in the capacitor is then
  dissipated in the resistor.

{Equation \eqref{eq:heat-func-alpha} tells us that} the price
to pay for an accelerated ($\alpha >1$)  charge or discharge is a
larger dissipation. {This equation} is also interesting when the process is decelerated, i.e., when $\alpha<1$. In this case, the heat decreases and vanishes in the limit as $\alpha\to 0^{+}$. We recover here the well-known result of a quasistatic transformation. The heat tends to zero, but the characteristic time required to charge the capacitor tends to infinity.

\begin{figure}
\centering
\includegraphics[width=8.3cm]{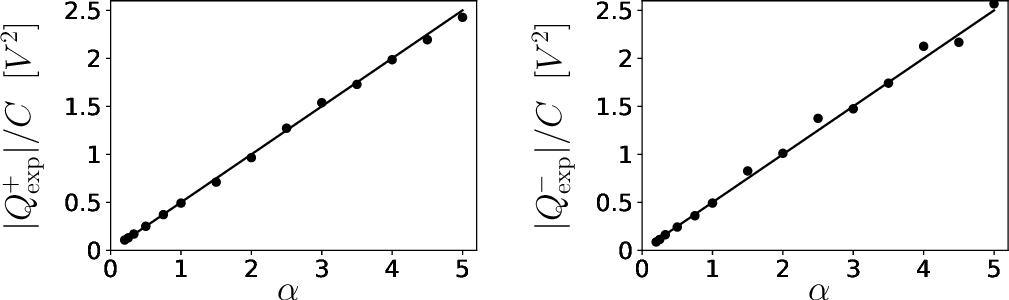}
\caption{Dissipated heat for the charging (left) and discharging
  (right) processes deduced from the data acquisition (see text).  The
  value for $\alpha=1$ is obtained for the reference step
  protocol. When the processes are accelerated (slowed down), the heat
  expense increases (decreases). We find an excellent agreement with
  the theoretical prediction (solid line) from
  Eq.~\eqref{eq:heat-func-alpha}. The {quasistatic} limit,
  $-Q \to 0^+$, is approached when the transformation is
  {infinitely slowed down, i.e.,} $\alpha \to 0^+$.}
\label{figenscaling}
\end{figure}

We have verified the result \eqref{eq:heat-func-alpha} in our circuit for both processes, charging and discharging. For this purpose, we collect the generated voltage $\varepsilon_{\rm exp}(t)$ and the capacitor voltage $\varepsilon^{\exp}_C(t)$ over the time interval of interest, for various values of $\alpha$. We then calculate the experimental value $Q^{\pm}_{\exp}$ of the heat
\begin{align}
-Q^{\pm}_{\exp}(\alpha) &= \int_0^{\infty} (\varepsilon(t)-\varepsilon_{C}(t))C \frac{d\varepsilon_{C}}{dt} dt\nonumber \\
&= C\sum_{n=0}^\infty (\varepsilon_{\rm exp}(n\Delta t) -\varepsilon^{\exp}_C( n\Delta t))
\nonumber\\
&\times \left[ \varepsilon^{\exp}_C( (n+1)\Delta t)- \varepsilon^{\exp}_C( n\Delta t)\right],
\end{align}
where we have employed  Eq.~\eqref{eq:kir} for
$\varepsilon(t)-\varepsilon_{C}(t)$, the definition of $Q$ in
Eq.~\eqref{eq:ener-heat}, and taken into account that $i=C d\varepsilon_C/dt=\varepsilon_R/R$. In Fig.~\ref{figenscaling}, we plot $-Q^{\pm}_{\exp} /C$ for
different scaling factors $\alpha$. The smallest  (largest) value for
the scaling factor is here $\alpha=0.2$ ($\alpha=5$), which  slows
down (speeds up)  the charging or discharging time by a factor of 5.

In Fig.~\ref{figenscaling}, we observe an excellent agreement within the accuracy of the measurement and the generation of the desired voltage. As expected, we measure a reduction (increase) of the energy dissipated in the resistor when we actively slow down (accelerate) the charging process. By increasing the charging time, we approach the {quasistatic} limit. The very same data have been acquired and analyzed for the discharging of the capacitor, evidencing the validity of Eq.~\eqref{eq:heat-func-alpha} for both kind of processes. 


\section{Energy optimization}\label{sec:energy-opt}

We have explained that dissipation is encapsulated in the heat term $-Q>0$, which depends on the entire charging time function, $q(t)$, i.e., it is a functional thereof. We can now ask ourselves the following question: how can we charge a capacitor to a given value $q_f$ in a finite amount of time $t_f$ with minimal dissipation?

To address this question, we write the dissipated heat in the charging process as the time integral, over the interval $[0,t_{f}]$ of the ``Lagrangian'' $\mathcal{L}=R \dot{q}^{2}$,
\begin{equation}
-Q^{+}[q] = \int_0^{t_f} {\cal L}(q,\dot q)dt = R \int_0^{t_f} \dot q^2 dt.
\end{equation}
We want to minimize the dissipated heat $-Q$ in the charging process, i.e., with the boundary conditions
\begin{equation}\label{eq:bc}
  q(0)=0, \quad q(t_{f})=q_{f}.
\end{equation}
The function $q(t)$ that minimizes $-Q$ is thus given by the solution of the Euler-Lagrange equation
\begin{equation}
\left.\frac{d}{dt} \left( \frac{\partial {\cal L}}{\partial \dot q} \right)-  \frac{\partial {\cal L}}{\partial q}\right|_{q_{\opt}}=0 \implies \ddot{q}_{opt}=0.
\end{equation}
Taking into account the boundary conditions \eqref{eq:bc}, we have:
\begin{equation}
q_{\opt}(t)=q_f \frac{t}{t_f}, \quad 0\le t\le t_{f}.
\end{equation}
The corresponding minimum dissipated heat is
  \begin{equation}
    -Q^{+}_{\opt}=R\int_{0}^{t_{f}} dt\, \dot{q}_{\opt}^{2}(t)=R \frac{q_{f}^{2}}{t_{f}}= \frac{q_{f}^{2}}{C}\frac{\tau}{t_{f}}=\frac{2\tau}{t_{f}}\Delta U^{+}.
  \end{equation}

\begin{figure}
\centering
\includegraphics[width=7cm]{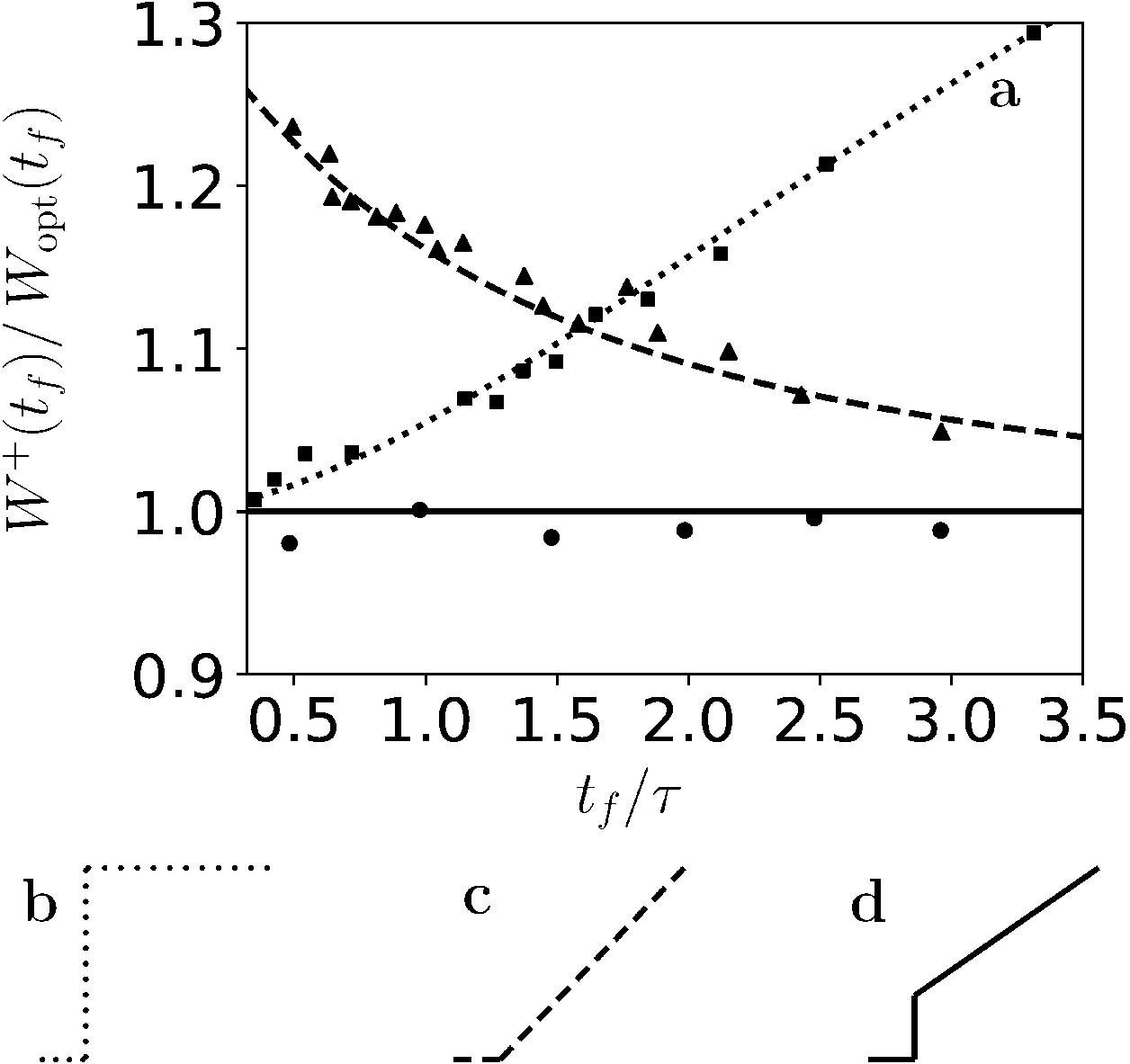}
\caption{Charge of a capacitor. \textbf{a}: Energy cost as a function
  of the final time $t_f$ to charge the capacitor to the voltage
  $\varepsilon_f=1$~V normalized to the minimal total energy cost in
  the finite time $t_f$. The different shapes used on the generator
  voltage are represented in \textbf{b}, \textbf{c} and
  \textbf{d}. The optimal driving voltage is schematically represented
  in \textbf{d}. We also represent the two limiting cases for the
  voltage driving: the step voltage (\textbf{b}) valid for $t_f\ll
  \tau$ and the linear ramp voltage (\textbf{c}) valid in the opposite
  limit. We have also plotted the results of our experimental data for
  the different driving without any adjustable parameters: optimal
  (disk), step shape (square), and linear shape (triangle).
}
\label{FigOptimisation}
\end{figure}

By inserting this relation in Eq.~\eqref{eqrc}, we infer the voltage to be applied for this optimal solution :
\begin{equation}\label{eq:epsilon-opt}
\varepsilon_{\opt}^{+}(t)=\varepsilon_f \left( \frac{\tau}{t_f} + \frac{t}{t_f} \right), \quad 0 < t < t_{f},
\end{equation}
recalling Eq.~\eqref{eq:qf-def}. The corresponding driving is nothing
but a linear variation of the voltage over an offset (see
Fig.~\ref{FigOptimisation}d). It must be remarked that, for $t>t_f$, it is necessary to provide the value of the voltage
associated to the final value of the charge, i.e.,
$\varepsilon_{\opt}(t)=\varepsilon_f$ for $t>t_f$, to keep the final
state stationary.

It is interesting to analyse the limit behaviour of the optimal
voltage in Eq.~\eqref{eq:epsilon-opt} for very short and very long
connection times. For short final time, $t_f\ll \tau$, the optimal
voltage adopts a step shape for $t<t_{f}$, as shown in
Fig.~\ref{FigOptimisation}b, with a constant value in this time
window that increases as $t_{f}^{-1}$. Therefore, in the limit
$t_{f}\to 0^{+}$, $\varepsilon_{\opt}^{+}(t)$ tends to a Dirac-delta
function. For long final time, $t_f \gg \tau$,
$\varepsilon_{\opt}^{+}(t)$ adopts to a linear shape, as shown in
Fig.~\ref{FigOptimisation}c. The energy input from the generator for
the optimal driving can be readily calculated:
\begin{equation} W_{\opt}^{+} (t_f)=\Delta
U^{+}-Q_{\opt}^{+}(t_{f})=\left( 1 + \frac{2\tau}{t_f} \right) \Delta
U^{+}.
\end{equation}
This value is used in the following to normalize the
data. It is therefore represented as a straight horizontal solid line
in Fig.~\ref{FigOptimisation}a.

It is worth noticing that we derive here a speed-limit inequality: {since}
\begin{equation}
|Q| \geq \left|Q_{\opt}(t_{f})\right|=\frac{2 \tau}{t_f}\Delta U^{+},
\end{equation}
{one concludes that}
\begin{equation}
  t_{f}|Q|\geq 2\tau \Delta U^{+},
\end{equation}
which gives us a trade-off between operation time and dissipation.
Analogous time-speed inequalities have been recently derived in
different contexts, both for quantum and classical
systems.\cite{aurell_optimal_2011,sivak_thermodynamic_2012,deffner_quantum_2017,okuyama_quantum_2018,shiraishi_speed_2018,shanahan_quantum_2018,funo_speed_2019,shiraishi_speed_2020,plata_finite-time_2020,deffner_quantum_2020,ito_stochastic_2020,van_vu_geometrical_2021,prados_optimizing_2021,lee_speed_2022,patron_thermal_2022,dechant_minimum_2022,guery-odelin_driving_2023}
The faster we want to charge the capacitor, the higher the energy cost (heat expense)
will be. Conversely, we identify how one approaches the
quasistatic limit with a minimal dissipation and for
very long, formally infinite, final time. 

For a constant voltage over the time interval $t_f$---step shape
represented in Fig.~\ref{FigOptimisation}b, the total energy input
from the generator is found to be equal to (see~Appendix~\ref{app:deriv-eqs})
\begin{equation}
W_{\rm ste} ^{+}(t_f)=\frac{2\Delta U^{+}}{1-e^{-t_f/\tau}}.
\label{eqwsq}
\end{equation}
The corresponding curve is represented as a dotted line in
Fig.~\ref{FigOptimisation}a. For the linear shape represented
schematically in Fig.~\ref{FigOptimisation}c, we obtain the following
total energy cost (see also~Appendix~\ref{app:deriv-eqs})
\begin{equation}
W_{\rm lin}^{+} (t_f)=2\Delta U^{+}\, \frac{(1+t_f/\tau)e^{-t_f/\tau} +t_f^2/2\tau^2-1}{(t_f/\tau-1+e^{-t_f/\tau})^2}.
\label{eqtrq}
\end{equation}

In Fig.~\ref{FigOptimisation}a, we observe an excellent agreement
between the data processed and the theoretical predictions detailed
above---the measured error is always within 2-3 \%. The experimental
values of the work are obtained by using the voltage data of the
generator and the capacitor through
\begin{align}
\label{eq:defWmes}
W^{+}_{\exp} &= \int_0^{\infty} \varepsilon(t)C \frac{d\varepsilon_{C}}{dt} dt\nonumber \\
&= C\sum_{n=0}^\infty \varepsilon_{\rm gen}(n\Delta t) \left[ \varepsilon^{\exp}_C( (n+1)\Delta t)- \varepsilon^{\exp}_C( n\Delta t)\right],
\end{align}
where we have used the definition of $W$ in Eq.~\eqref{eq:defW} and
taken into account that $i=C d\varepsilon_C/dt$. Note that the upper
limit in Eq.~\eqref{eq:defWmes} is infinite, instead of $t_f$, without
loss of generality, since for $t>t_f$ the process is stationary and
there is no energy exchange.  We clearly observe the crossover between
the two asymptotic shapes of the optimal driving function. As we plot
the ratio $W^{+}(t_{f})/W_{\opt}^{+}(t_{f})$, we directly deduce that
the best choice for the driving voltage can reduce the energy
consumption up to about $ 25 \%$ in the range of final times
considered here.


\section{Discussion}\label{sec:discussion}

In this article, we have investigated how to control a
  voltage generator connected to an RC series electrical circuit to
  accelerate or decelerate the charging or discharging of the
  capacitor. The active discharge under consideration here
    differs from active discharge involving only passive electric
    dipoles, where, for example, a switch is deliberately or actively
    closed to enable the capacitor to quickly discharge through
    another resistor with small resistance. This simple problem lends
  itself very well to a thermodynamic analysis using the first
  law, where internal energy, work, and heat can be clearly
  identified. Internal energy is associated with the energy stored in
  the capacitor, whereas work and heat are associated with the energy
  input from the battery and the energy dissipated in the resistor,
  respectively: concepts that are all familiar to undergraduate
  students.

First, we have followed an inverse engineering approach with
  time-rescaled solutions, and also analyzed the energy cost, in terms
  of the dissipated Joule heat, of such controls. Specifically, the
  idea of an infinitely slow quasistatic transformation, with zero
  dissipation, naturally appears. Second, we have investigated the
  problem of finding the protocol that minimizes the energy cost for a
  given charging time.  The minimization of the dissipated heat
  provides an opportunity to motivate and show the power of calculus
  of variations and optimal control theory with a simple but
  illuminating example, by combining concepts of different fields that
  are usually taught separately.  In addition, it facilitates our
  introducing undergraduate students to advanced concepts of current
  research, such as the emergence of speed limits, i.e., the
  inevitable trade-off between time operation and dissipation.

Our problem of minimal energy cost in an RC circuit
generalizes to formally equivalent problems, also accessible
  to undergraduate students. Take, for instance, the case of an
overdamped particle in one dimension submitted to a conservative
harmonic force ${F}_h=- k {x}$ and a non-conservative friction force
${F}_{v}=-\gamma {v}$. We add an external force ${F} (t)$ with the
objective of driving the elongation ${x}(t)$ from ${x}(0)={0}$ to a
given final value ${x}(t_f)={x}_f$ in a finite amount of time, while
dissipating as little energy as possible. In this equivalent
mechanical problem, the internal energy is $U=k x^{2}/2$, the work is
$\int_{0}^{t_{f}}F(t)v(t)dt$, and the dissipated heat is
$Q=-\int_{0}^{t_{f}} \gamma v^{2}dt$.

One possible extension is the consideration of an RLC circuit---or,
equivalently, an underdamped particle of mass $m$, also under the
action of the harmonic ${F}_{h}$, friction ${F}_{v}$, and external
${F}$ forces described above. Note that the inductance $L$ modifies
the evolution equation to be a second-order differential equation, and
thus both $q(t)$ and $i(t)$ must be continuous functions of
time. Moreover, the control problem for the charging process is
modified, now we would like to reach a final charge $q(t_{f})=q_{f}$
with null current $i(t_{f})=i_{f}=0$ from the initial state in which
the capacitor is discharged and there is no current, $q(0)=0$,
$i(0)=0$. Still, heat is also given by the Joule contribution and, as
a result, minimizing the dissipated heat yields the same equation
$\ddot q=0$, or $ di/dt=0$. This entails that the optimal current must
be $i(t)=q_{f}/t_{f}$ in order to reproduce the target charge, but
this does not meet the boundary conditions for the
current. Mathematically, this difficulty may be solved by introducing
finite jumps in the current at the initial and final times, i.e.,
$i(0)=i(t_{f})=0$ but $i(t)=q_{f}/t_{f}$ for
$0^{+}<t<t_{f}^{-}$.\footnote{These finite jump discontinuities do not
  contribute to the dissipated heat but make the source voltage have
  two delta peaks at the initial and final times, diverging to
  $+\infty$ and $-\infty$, respectively. This behavior is analogous to
  that found in the minimization of the irreversible work of an
  underdamped particle under harmonic confinement in a thermal bath at
  temperature $T$.\cite{gomez-marin_optimal_2008} }

Optimization of energy resources is a stimulating topic for
  students, since it undoubtedly constitutes one of the main
  challenges for the next generations of physicists. In fact,
  capacitors play a crucial role in renewable energy systems by
helping to store and manage the fluctuating energy produced by sources
like wind and solar power, specifically by storing excess energy
during periods of high production and discharging it when needed. In
solar systems, they are employed in photovoltaic inverters that
typically make extensive use of large-sized capacitors to store
electric energy. Capacitors are also a key ingredient in the
electrical modeling of batteries, particularly in modeling the
behavior and characteristics of rechargeable batteries, such as
Lithium-ion batteries commonly used in electric
vehicles.\cite{hou_adaptive_2019,dreyer_hysteresis_2011,dreyer_behavior_2011,dreyer_thermodynamic_2010,li_30_2018,wang_comprehensive_2020,manthiram_reflection_2020}
Finally, we would like to highlight that the problem
  investigated in this paper is also easily dealt with as a computer
interface project, providing the interested instructor with a
  comprehensive problem comprising all the ingredients of a research
activity: modeling, implementation, and data processing to access the
physical quantities of interest.

\appendix

\section{General charging/discharging process}
\label{app:partial-charg-discharg}

In the main text, we have discussed in detail the charging and
discharging process when either the initial or final value of the
charge stored in the capacitor was zero. In this appendix, we refer to
these protocols as ``pure'' charging or discharging processes. For the
sake of completeness, we provide here the general theory when
arbitrary initial and final values for the charge are considered,
allowing us to study, for example, partial charging or discharging
processes.

Let us start by introducing the reference step protocol in this
general case. We assume the capacitor has a certain charge state
$q(0)=q_i$ at the initial time $t=0$, associated with the
corresponding value of the voltage $\varepsilon_i=q_i/C$. At
$t=0^{+}$, the voltage in the source is abruptly changed to
$\varepsilon_f$ and the charge in the capacitor evolves following
Eq.~\eqref{eqrc} towards the final value $q(\infty)=q_f$, where
$q_{f}$ is given by Eq.~\eqref{eq:qf-def}. Hence, the reference
solution is now
\begin{equation}
q_r(t)=C \left[\varepsilon_f + \left(\varepsilon_i-\varepsilon_f \right) e^{-t/\tau} \right],
\end{equation}
which is consistent with the ``pure'' case  of charging or discharging
processes, by either taking $\varepsilon_i \to 0$ or
$\varepsilon_f \to 0$, respectively. \footnote{Note that
  the notation employed here is slightly different from that of the
  main text: here, $\varepsilon_{i}$ and $\varepsilon_{f}$ always
  stand for the initial and final voltage of the corresponding
  process, whereas in the main text we only introduced $\epsilon_{f}$
  as the final voltage of the charging process, which was the initial
  one for the subsequent discharging process.}

The energy balance is readily obtained by direct application of the
definitions of work, internal energy, and heat. Using
Eqs.~\eqref{eq:first-principle} and \eqref{eq:ener-heat}, we get
\begin{subequations}
\begin{align}
W_r &= C \varepsilon_f \left( \varepsilon_f - \varepsilon_i \right), \\
\Delta U & = \frac{1}{2} C\left( \varepsilon_f^2 - \varepsilon_i^2 \right) ,\\
-Q_r &=\frac{1}{2} C \left( \varepsilon_f - \varepsilon_i \right)^2.\label{eq:Qr-gen}
\end{align}
\end{subequations}
Once more, ``pure'' charging or discharging processes are reobtained
when taking $\varepsilon_i \to 0$ or $\varepsilon_f \to 0$,
respectively. As already described in the main text, (i) ``pure''
charging process involves that half of the input work is invested in
changing the energy while the other half is dissipated,
$W^+_r=2 \Delta U=-2Q^+_r>0$ and (ii) ``pure'' discharging process
involve no work, since there is no power input and the change of
energy is directly due to dissipation, $\Delta U^{-}=Q_r^{-}<0$. In
these ``pure'' charging/discharging processes, we have
$\Delta U^{\pm} = \mp Q_r^{\pm}$, since either $\varepsilon_i \to 0$
or $\varepsilon_f \to 0$. This is no longer the case when considering
general partial charging or discharging processes.

In the main text, all energetic quantities were always found to be
proportional to $\Delta U$, but this simple scaling does not hold when
considering general charging/discharging processes. However, there is
a remarkable scaling property that persists for general charging or
discharging processes, which applies to the dissipated heat in any of
the considered transformations. In all cases (reference protocol, inverse engineered protocol, optimal protocol, step and linear
protocols), the heat is proportional to
$\left(\varepsilon_f - \varepsilon_i \right)^2$, as the
reference heat in Eq.~\eqref{eq:Qr-gen} is. Specifically, it can be
proven that
\begin{subequations}
\begin{align}
Q(\alpha)&=\alpha Q_r,\\
Q_{\opt}(t_f)&=\frac{2 \tau}{t_f} Q_r,\\
Q_{\rm ste}(t_f)&=\coth\!\left(\frac{t_{f}}{2\tau}\right) Q_r,\\
Q_{\rm lin}(t_f)&=\frac{2t_f/ \tau -3+4e^{-t_f/ \tau }-e^{-2t_f/ \tau
                 }}
                 {\left[t_f/ \tau -1+e^{-t_f/ \tau }\right]^{2}}Q_r.
\end{align}
\end{subequations}
In the above equations, $Q(\alpha)$ is the heat for the inverse
engineered protocol with speed-up factor $\alpha$, $Q_{\opt}(t_{f})$
is the heat for the optimal protocol, which still implies
$\ddot{q}=0$, with duration $t_{f}$, $Q_{\rm ste}(t_f)$ is the heat for
the square protocol, and $Q_{\rm lin}(t_f)$ is the heat for the
triangular protocol. The latter two are naturally consistent with the
expressions in Eqs.~\eqref{eqwsq} and \eqref{eqtrq} of the main text,
which are derived in Appendix~\ref{app:deriv-eqs}.  Of course, when
$\varepsilon_i \to 0$ or $\varepsilon_f \to 0$, $|\Delta U |= |Q_r|$
and the universal scaling for the heat is inherited by the work, as
shown in the main text.


\section{Derivation of Eqs.~\eqref{eqwsq} and \eqref{eqtrq}}\label{app:deriv-eqs}

In this appendix, we detail the calculation of the expressions for
$W_{\rm ste}^{+} (t_f)$ and $W_{\rm lin}^{+} (t_f)$ that are given in
the main text. 

First, we consider the charging under a constant voltage
  $\bar{\varepsilon}$. The charge evolves according to
Eq.~\eqref{eq:q(t)-ref-proc}, with the change
$\varepsilon_{f}\to\varepsilon_{\rm ste}=\bar{\varepsilon}$ and the current as
\begin{equation}
i_{\rm ste}(t)=\frac{\bar{\varepsilon}}{R}e^{-t/\tau}.
\end{equation}
We infer the expression for the work
\begin{equation}
W_{\rm ste}^{+} (t_f) = \varepsilon_{\rm ste} \int_0^{t_f} i_{\rm ste}(t) dt = C \bar{\varepsilon}^2 (1-e^{-t_f/\tau}).
\label{wsq1}
\end{equation}
To express $W_{\rm ste}^{+} (t_f)$ in terms of $\Delta U^+=C\varepsilon_f^2/2$, we use 
\begin{equation}
\varepsilon_f=\varepsilon(t_f)= \frac{q(t_f)}{C}=\bar{\varepsilon} (1-e^{-t_f/\tau}).
\label{wsq2}
\end{equation}
Combining Eqs.~\eqref{wsq1} and \eqref{wsq2}, we recover the result for the work $W_{\rm ste} ^{+}(t_f)$ of Eq.~\eqref{eqwsq}. 

For the linear variation of the generator voltage, we obtain the charge as a function of time by solving the differential equation
\begin{equation}
R \, \dot q_{\rm lin} + \frac{q_{\rm lin}}{C}=\varepsilon_{\rm lin}(t)=\hat{\varepsilon} \frac{t}{t_f},
\end{equation}
with $\hat{\varepsilon}/t_f$ being the constant slope of the linear variation of the voltage source. Using the initial condition $q(0)=0$, we find 
\begin{equation}
q_{\rm lin}(t)=\frac{\hat{\varepsilon} \tau^2}{Rt_f} \left[ e^{-t/\tau}-1+\frac{t}{\tau}   \right]  
\end{equation}
for the charge, and 
\begin{equation}
i_{\rm lin}(t)= \frac{d q_{\rm lin}(t)}{dt}=\frac{\hat{\varepsilon} \tau}{Rt_f} (1-e^{-t/\tau}).
\end{equation}
for the current. Now, the work $W_{\rm lin} ^{+}(t_f)$ is readily calculated:
\begin{align}
W_{\rm lin} ^{+}(t_f)  & =   \int_0^{t_f}  \varepsilon_{\rm lin}(t) i_{\rm lin}(t) dt  \nonumber \\ 
& =  C\hat{\varepsilon}^2 \frac{\tau^2}{t_f^2} \left[    \frac{1}{2}\left( \frac{t_f}{\tau}\right)^2-1 + \left(1+ \frac{t_f}{\tau}\right) e^{-t_f/\tau}\right].
\label{wtr1}
\end{align}
Once again, to express this result in terms of $\Delta U^+=C\varepsilon_f^2/2$, we rely on the relation
\begin{equation}
\varepsilon_f=\varepsilon(t_f)= \frac{q(t_f)}{C}=\frac{\hat{\varepsilon} \tau}{t_f} \left[ e^{-t_f/\tau}-1+\frac{t_f}{\tau}   \right].   
\label{wtr2}
\end{equation}
Combining Eqs.~\eqref{wtr1} and \eqref{wtr2}, we recover the result for the work $W_{\rm lin} ^{+}(t_f)$ of Eq.~\eqref{eqtrq}.

\begin{acknowledgments}
 CAP and AP acknowledge financial support from Grant
 PID2021-122588NB-I00 funded by MCIN/AEI/10.13039/501100011033/ and
  by ``ERDF A way of making Europe''. CAP, AP, and DGO also
 acknowledge financial support from Grant ProyExcel\_00796 funded by
 Junta de Andalucía's PAIDI 2020 program. CAP acknowledges the
 funding received from European Union’s Horizon Europe–Marie
 Skłodowska-Curie 2021 program through the Postdoctoral Fellowship
 with Reference 101065902 (ORION).
\end{acknowledgments}

\bibliography{Mi-biblioteca-13-jun-2024}

\begin{thebibliography}{41}%
\makeatletter
\providecommand \@ifxundefined [1]{%
 \@ifx{#1\undefined}
}%
\providecommand \@ifnum [1]{%
 \ifnum #1\expandafter \@firstoftwo
 \else \expandafter \@secondoftwo
 \fi
}%
\providecommand \@ifx [1]{%
 \ifx #1\expandafter \@firstoftwo
 \else \expandafter \@secondoftwo
 \fi
}%
\providecommand \natexlab [1]{#1}%
\providecommand \enquote  [1]{``#1''}%
\providecommand \bibnamefont  [1]{#1}%
\providecommand \bibfnamefont [1]{#1}%
\providecommand \citenamefont [1]{#1}%
\providecommand \href@noop [0]{\@secondoftwo}%
\providecommand \href [0]{\begingroup \@sanitize@url \@href}%
\providecommand \@href[1]{\@@startlink{#1}\@@href}%
\providecommand \@@href[1]{\endgroup#1\@@endlink}%
\providecommand \@sanitize@url [0]{\catcode `\\12\catcode `\$12\catcode
  `\&12\catcode `\#12\catcode `\^12\catcode `\_12\catcode `\%12\relax}%
\providecommand \@@startlink[1]{}%
\providecommand \@@endlink[0]{}%
\providecommand \url  [0]{\begingroup\@sanitize@url \@url }%
\providecommand \@url [1]{\endgroup\@href {#1}{\urlprefix }}%
\providecommand \urlprefix  [0]{URL }%
\providecommand \Eprint [0]{\href }%
\providecommand \doibase [0]{https://doi.org/}%
\providecommand \selectlanguage [0]{\@gobble}%
\providecommand \bibinfo  [0]{\@secondoftwo}%
\providecommand \bibfield  [0]{\@secondoftwo}%
\providecommand \translation [1]{[#1]}%
\providecommand \BibitemOpen [0]{}%
\providecommand \bibitemStop [0]{}%
\providecommand \bibitemNoStop [0]{.\EOS\space}%
\providecommand \EOS [0]{\spacefactor3000\relax}%
\providecommand \BibitemShut  [1]{\csname bibitem#1\endcsname}%
\let\auto@bib@innerbib\@empty
\bibitem [{\citenamefont {Callen}(1985)}]{callen_thermodynamics_1985}%
  \BibitemOpen
  \bibfield  {author} {\bibinfo {author} {\bibfnamefont {H.}~\bibnamefont
  {Callen}},\ }\href@noop {} {\emph {\bibinfo {title} {Thermodynamics and an
  {Introduction} to {Thermostatistics}}}}\ (\bibinfo  {publisher} {Wiley},\
  \bibinfo {year} {1985})\BibitemShut {NoStop}%
\bibitem [{\citenamefont {Goldstein}(1980)}]{goldstein_mechanics_1980}%
  \BibitemOpen
  \bibfield  {author} {\bibinfo {author} {\bibfnamefont {H.}~\bibnamefont
  {Goldstein}},\ }\href@noop {} {\emph {\bibinfo {title} {Classical
  mechanics}}},\ \bibinfo {edition} {2nd}\ ed.,\ Addison-Wesley series in
  physics\ (\bibinfo  {publisher} {Addison-Wesley},\ \bibinfo {year}
  {1980})\BibitemShut {NoStop}%
\bibitem [{\citenamefont {Landau}\ and\ \citenamefont
  {Lifshitz}(1976)}]{landau_mechanics_1976}%
  \BibitemOpen
  \bibfield  {author} {\bibinfo {author} {\bibfnamefont {L.~D.}\ \bibnamefont
  {Landau}}\ and\ \bibinfo {author} {\bibfnamefont {E.~M.}\ \bibnamefont
  {Lifshitz}},\ }\href@noop {} {\emph {\bibinfo {title} {Course of Theoretical
  Physics: Vol. 1, Mechanics}}},\ \bibinfo {edition} {3rd}\ ed.,\ Course of
  Theoretical Physics\ (\bibinfo  {publisher} {Butterworth-Heinemann},\
  \bibinfo {year} {1976})\BibitemShut {NoStop}%
\bibitem [{\citenamefont {Sekimoto}(2010)}]{sekimoto_stochastic_2010}%
  \BibitemOpen
  \bibfield  {author} {\bibinfo {author} {\bibfnamefont {K.}~\bibnamefont
  {Sekimoto}},\ }\href@noop {} {\emph {\bibinfo {title} {Stochastic
  {Energetics}}}}\ (\bibinfo  {publisher} {Springer},\ \bibinfo {year}
  {2010})\BibitemShut {NoStop}%
\bibitem [{\citenamefont {Guéry-Odelin}\ \emph {et~al.}(2023)\citenamefont
  {Guéry-Odelin}, \citenamefont {Jarzynski}, \citenamefont {Plata},
  \citenamefont {Prados},\ and\ \citenamefont
  {Trizac}}]{guery-odelin_driving_2023}%
  \BibitemOpen
  \bibfield  {author} {\bibinfo {author} {\bibfnamefont {D.}~\bibnamefont
  {Guéry-Odelin}}, \bibinfo {author} {\bibfnamefont {C.}~\bibnamefont
  {Jarzynski}}, \bibinfo {author} {\bibfnamefont {C.~A.}\ \bibnamefont
  {Plata}}, \bibinfo {author} {\bibfnamefont {A.}~\bibnamefont {Prados}},\ and\
  \bibinfo {author} {\bibfnamefont {E.}~\bibnamefont {Trizac}},\ }\bibfield
  {title} {\enquote {\bibinfo {title} {Driving rapidly while remaining in
  control: classical shortcuts from {Hamiltonian} to stochastic dynamics},}\
  }\href {https://doi.org/10.1088/1361-6633/acacad} {\bibfield  {journal}
  {\bibinfo  {journal} {Reports on Progress in Physics}\ }\textbf {\bibinfo
  {volume} {86}},\ \bibinfo {pages} {035902} (\bibinfo {year}
  {2023})}\BibitemShut {NoStop}%
\bibitem [{\citenamefont {Blaber}\ and\ \citenamefont
  {Sivak}(2023)}]{blaber_optimal_2023}%
  \BibitemOpen
  \bibfield  {author} {\bibinfo {author} {\bibfnamefont {S.}~\bibnamefont
  {Blaber}}\ and\ \bibinfo {author} {\bibfnamefont {D.~A.}\ \bibnamefont
  {Sivak}},\ }\bibfield  {title} {\enquote {\bibinfo {title} {Optimal control
  in stochastic thermodynamics},}\ }\href
  {https://doi.org/10.1088/2399-6528/acbf04} {\bibfield  {journal} {\bibinfo
  {journal} {Journal of Physics Communications}\ }\textbf {\bibinfo {volume}
  {7}},\ \bibinfo {pages} {033001} (\bibinfo {year} {2023})}\BibitemShut
  {NoStop}%
\bibitem [{\citenamefont {Liberzon}(2012)}]{liberzon_calculus_2012}%
  \BibitemOpen
  \bibfield  {author} {\bibinfo {author} {\bibfnamefont {D.}~\bibnamefont
  {Liberzon}},\ }\href@noop {} {\emph {\bibinfo {title} {Calculus of
  {Variations} and {Optimal} {Control} {Theory}: {A} {Concise}
  {Introduction}}}}\ (\bibinfo  {publisher} {Princeton University Press},\
  \bibinfo {year} {2012})\BibitemShut {NoStop}%
\bibitem [{\citenamefont {Schmiedl}\ and\ \citenamefont
  {Seifert}(2007)}]{schmiedl_optimal_2007}%
  \BibitemOpen
  \bibfield  {author} {\bibinfo {author} {\bibfnamefont {T.}~\bibnamefont
  {Schmiedl}}\ and\ \bibinfo {author} {\bibfnamefont {U.}~\bibnamefont
  {Seifert}},\ }\bibfield  {title} {\enquote {\bibinfo {title} {Optimal
  {Finite}-{Time} {Processes} {In} {Stochastic} {Thermodynamics}},}\ }\href
  {https://doi.org/10.1103/PhysRevLett.98.108301} {\bibfield  {journal}
  {\bibinfo  {journal} {Physical Review Letters}\ }\textbf {\bibinfo {volume}
  {98}},\ \bibinfo {pages} {108301} (\bibinfo {year} {2007})}\BibitemShut
  {NoStop}%
\bibitem [{\citenamefont {Aurell}, \citenamefont {Mejía-Monasterio},\ and\
  \citenamefont {Muratore-Ginanneschi}(2011)}]{aurell_optimal_2011}%
  \BibitemOpen
  \bibfield  {author} {\bibinfo {author} {\bibfnamefont {E.}~\bibnamefont
  {Aurell}}, \bibinfo {author} {\bibfnamefont {C.}~\bibnamefont
  {Mejía-Monasterio}},\ and\ \bibinfo {author} {\bibfnamefont
  {P.}~\bibnamefont {Muratore-Ginanneschi}},\ }\bibfield  {title} {\enquote
  {\bibinfo {title} {Optimal {Protocols} and {Optimal} {Transport} in
  {Stochastic} {Thermodynamics}},}\ }\href
  {https://doi.org/10.1103/PhysRevLett.106.250601} {\bibfield  {journal}
  {\bibinfo  {journal} {Physical Review Letters}\ }\textbf {\bibinfo {volume}
  {106}},\ \bibinfo {pages} {250601} (\bibinfo {year} {2011})}\BibitemShut
  {NoStop}%
\bibitem [{\citenamefont {Sivak}\ and\ \citenamefont
  {Crooks}(2012)}]{sivak_thermodynamic_2012}%
  \BibitemOpen
  \bibfield  {author} {\bibinfo {author} {\bibfnamefont {D.~A.}\ \bibnamefont
  {Sivak}}\ and\ \bibinfo {author} {\bibfnamefont {G.~E.}\ \bibnamefont
  {Crooks}},\ }\bibfield  {title} {\enquote {\bibinfo {title} {Thermodynamic
  {Metrics} and {Optimal} {Paths}},}\ }\href
  {https://doi.org/10.1103/PhysRevLett.108.190602} {\bibfield  {journal}
  {\bibinfo  {journal} {Physical Review Letters}\ }\textbf {\bibinfo {volume}
  {108}},\ \bibinfo {pages} {190602} (\bibinfo {year} {2012})}\BibitemShut
  {NoStop}%
\bibitem [{\citenamefont {Masuda}\ and\ \citenamefont
  {Nakamura}(2010)}]{masuda_fast-forward_2010}%
  \BibitemOpen
  \bibfield  {author} {\bibinfo {author} {\bibfnamefont {S.}~\bibnamefont
  {Masuda}}\ and\ \bibinfo {author} {\bibfnamefont {K.}~\bibnamefont
  {Nakamura}},\ }\bibfield  {title} {\enquote {\bibinfo {title} {Fast-forward
  of adiabatic dynamics in quantum mechanics},}\ }\href
  {https://doi.org/10.1098/rspa.2009.0446} {\bibfield  {journal} {\bibinfo
  {journal} {Proceedings of the Royal Society A: Mathematical, Physical and
  Engineering Sciences}\ }\textbf {\bibinfo {volume} {466}},\ \bibinfo {pages}
  {1135--1154} (\bibinfo {year} {2010})}\BibitemShut {NoStop}%
\bibitem [{\citenamefont {Guéry-Odelin}\ \emph {et~al.}(2019)\citenamefont
  {Guéry-Odelin}, \citenamefont {Ruschhaupt}, \citenamefont {Kiely},
  \citenamefont {Torrontegui}, \citenamefont {Martínez-Garaot},\ and\
  \citenamefont {Muga}}]{guery-odelin_shortcuts_2019}%
  \BibitemOpen
  \bibfield  {author} {\bibinfo {author} {\bibfnamefont {D.}~\bibnamefont
  {Guéry-Odelin}}, \bibinfo {author} {\bibfnamefont {A.}~\bibnamefont
  {Ruschhaupt}}, \bibinfo {author} {\bibfnamefont {A.}~\bibnamefont {Kiely}},
  \bibinfo {author} {\bibfnamefont {E.}~\bibnamefont {Torrontegui}}, \bibinfo
  {author} {\bibfnamefont {S.}~\bibnamefont {Martínez-Garaot}},\ and\ \bibinfo
  {author} {\bibfnamefont {J.~G.}\ \bibnamefont {Muga}},\ }\bibfield  {title}
  {\enquote {\bibinfo {title} {Shortcuts to adiabaticity: {Concepts}, methods,
  and applications},}\ }\href {https://doi.org/10.1103/RevModPhys.91.045001}
  {\bibfield  {journal} {\bibinfo  {journal} {Reviews of Modern Physics}\
  }\textbf {\bibinfo {volume} {91}},\ \bibinfo {pages} {045001} (\bibinfo
  {year} {2019})}\BibitemShut {NoStop}%
\bibitem [{\citenamefont {Nakamura}, \citenamefont {Matrasulov},\ and\
  \citenamefont {Izumida}(2020)}]{nakamura_fast_2020}%
  \BibitemOpen
  \bibfield  {author} {\bibinfo {author} {\bibfnamefont {K.}~\bibnamefont
  {Nakamura}}, \bibinfo {author} {\bibfnamefont {J.}~\bibnamefont
  {Matrasulov}},\ and\ \bibinfo {author} {\bibfnamefont {Y.}~\bibnamefont
  {Izumida}},\ }\bibfield  {title} {\enquote {\bibinfo {title} {Fast-forward
  approach to stochastic heat engine},}\ }\href
  {https://doi.org/10.1103/PhysRevE.102.012129} {\bibfield  {journal} {\bibinfo
   {journal} {Physical Review E}\ }\textbf {\bibinfo {volume} {102}},\ \bibinfo
  {pages} {012129} (\bibinfo {year} {2020})}\BibitemShut {NoStop}%
\bibitem [{\citenamefont {Plata}\ \emph {et~al.}(2021)\citenamefont {Plata},
  \citenamefont {Prados}, \citenamefont {Trizac},\ and\ \citenamefont
  {Gu{\'e}ry-Odelin}}]{plata_taming_2021}%
  \BibitemOpen
  \bibfield  {author} {\bibinfo {author} {\bibfnamefont {C.~A.}\ \bibnamefont
  {Plata}}, \bibinfo {author} {\bibfnamefont {A.}~\bibnamefont {Prados}},
  \bibinfo {author} {\bibfnamefont {E.}~\bibnamefont {Trizac}},\ and\ \bibinfo
  {author} {\bibfnamefont {D.}~\bibnamefont {Gu{\'e}ry-Odelin}},\ }\bibfield
  {title} {\enquote {\bibinfo {title} {Taming the {Time} {Evolution} in
  {Overdamped} {Systems}: {Shortcuts} {Elaborated} from {Fast}-{Forward} and
  {Time}-{Reversed} {Protocols}},}\ }\href
  {https://doi.org/10.1103/PhysRevLett.127.190605} {\bibfield  {journal}
  {\bibinfo  {journal} {Physical Review Letters}\ }\textbf {\bibinfo {volume}
  {127}},\ \bibinfo {pages} {190605} (\bibinfo {year} {2021})}\BibitemShut
  {NoStop}%
\bibitem [{\citenamefont {Impens}\ \emph {et~al.}(2021)\citenamefont {Impens},
  \citenamefont {D'Angelis}, \citenamefont {Pinheiro},\ and\ \citenamefont
  {Guéry-Odelin}}]{impens_time_2021}%
  \BibitemOpen
  \bibfield  {author} {\bibinfo {author} {\bibfnamefont {F.}~\bibnamefont
  {Impens}}, \bibinfo {author} {\bibfnamefont {F.~M.}\ \bibnamefont
  {D'Angelis}}, \bibinfo {author} {\bibfnamefont {F.~A.}\ \bibnamefont
  {Pinheiro}},\ and\ \bibinfo {author} {\bibfnamefont {D.}~\bibnamefont
  {Guéry-Odelin}},\ }\bibfield  {title} {\enquote {\bibinfo {title} {Time
  scaling and quantum speed limit in non-{Hermitian} {Hamiltonians}},}\ }\href
  {https://doi.org/10.1103/PhysRevA.104.052620} {\bibfield  {journal} {\bibinfo
   {journal} {Physical Review A}\ }\textbf {\bibinfo {volume} {104}},\ \bibinfo
  {pages} {052620} (\bibinfo {year} {2021})}\BibitemShut {NoStop}%
\bibitem [{\citenamefont {Faure}\ \emph {et~al.}(2019)\citenamefont {Faure},
  \citenamefont {Ciliberto}, \citenamefont {Trizac},\ and\ \citenamefont
  {Guéry-Odelin}}]{faure_shortcut_2019}%
  \BibitemOpen
  \bibfield  {author} {\bibinfo {author} {\bibfnamefont {S.}~\bibnamefont
  {Faure}}, \bibinfo {author} {\bibfnamefont {S.}~\bibnamefont {Ciliberto}},
  \bibinfo {author} {\bibfnamefont {E.}~\bibnamefont {Trizac}},\ and\ \bibinfo
  {author} {\bibfnamefont {D.}~\bibnamefont {Guéry-Odelin}},\ }\bibfield
  {title} {\enquote {\bibinfo {title} {Shortcut to stationary regimes: {A}
  simple experimental demonstration},}\ }\href
  {https://doi.org/10.1119/1.5082933} {\bibfield  {journal} {\bibinfo
  {journal} {American Journal of Physics}\ }\textbf {\bibinfo {volume} {87}},\
  \bibinfo {pages} {125--129} (\bibinfo {year} {2019})}\BibitemShut {NoStop}%
\bibitem [{Note1()}]{Note1}%
  \BibitemOpen
  \bibinfo {note} {Note that we have written the first law as $\Delta U = Q+W$,
  i.e. we follow the usual sign convention in which energy exchanges, in the
  form of either heat or work, increasing (decreasing) the internal energy of
  the system are positive (negative). Hence, dissipation involves a negative
  $Q$, which motivates our writing of the non-negative dissipated heat as
  $-Q$.}\BibitemShut {Stop}%
\bibitem [{\citenamefont {Deffner}\ and\ \citenamefont
  {Campbell}(2017)}]{deffner_quantum_2017}%
  \BibitemOpen
  \bibfield  {author} {\bibinfo {author} {\bibfnamefont {S.}~\bibnamefont
  {Deffner}}\ and\ \bibinfo {author} {\bibfnamefont {S.}~\bibnamefont
  {Campbell}},\ }\bibfield  {title} {\enquote {\bibinfo {title} {Quantum speed
  limits: from {Heisenberg}’s uncertainty principle to optimal quantum
  control},}\ }\href {https://doi.org/10.1088/1751-8121/aa86c6} {\bibfield
  {journal} {\bibinfo  {journal} {Journal of Physics A: Mathematical and
  Theoretical}\ }\textbf {\bibinfo {volume} {50}},\ \bibinfo {pages} {453001}
  (\bibinfo {year} {2017})}\BibitemShut {NoStop}%
\bibitem [{\citenamefont {Okuyama}\ and\ \citenamefont
  {Ohzeki}(2018)}]{okuyama_quantum_2018}%
  \BibitemOpen
  \bibfield  {author} {\bibinfo {author} {\bibfnamefont {M.}~\bibnamefont
  {Okuyama}}\ and\ \bibinfo {author} {\bibfnamefont {M.}~\bibnamefont
  {Ohzeki}},\ }\bibfield  {title} {\enquote {\bibinfo {title} {Quantum {Speed}
  {Limit} is {Not} {Quantum}},}\ }\href@noop {} {\bibfield  {journal} {\bibinfo
   {journal} {Physical Review Letters}\ }\textbf {\bibinfo {volume} {120}},\
  \bibinfo {pages} {070402} (\bibinfo {year} {2018})}\BibitemShut {NoStop}%
\bibitem [{\citenamefont {Shiraishi}, \citenamefont {Funo},\ and\ \citenamefont
  {Saito}(2018)}]{shiraishi_speed_2018}%
  \BibitemOpen
  \bibfield  {author} {\bibinfo {author} {\bibfnamefont {N.}~\bibnamefont
  {Shiraishi}}, \bibinfo {author} {\bibfnamefont {K.}~\bibnamefont {Funo}},\
  and\ \bibinfo {author} {\bibfnamefont {K.}~\bibnamefont {Saito}},\ }\bibfield
   {title} {\enquote {\bibinfo {title} {Speed {Limit} for {Classical}
  {Stochastic} {Processes}},}\ }\href@noop {} {\bibfield  {journal} {\bibinfo
  {journal} {Physical Review Letters}\ }\textbf {\bibinfo {volume} {121}},\
  \bibinfo {pages} {070601} (\bibinfo {year} {2018})}\BibitemShut {NoStop}%
\bibitem [{\citenamefont {Shanahan}\ \emph {et~al.}(2018)\citenamefont
  {Shanahan}, \citenamefont {Chenu}, \citenamefont {Margolus},\ and\
  \citenamefont {del Campo}}]{shanahan_quantum_2018}%
  \BibitemOpen
  \bibfield  {author} {\bibinfo {author} {\bibfnamefont {B.}~\bibnamefont
  {Shanahan}}, \bibinfo {author} {\bibfnamefont {A.}~\bibnamefont {Chenu}},
  \bibinfo {author} {\bibfnamefont {N.}~\bibnamefont {Margolus}},\ and\
  \bibinfo {author} {\bibfnamefont {A.}~\bibnamefont {del Campo}},\ }\bibfield
  {title} {\enquote {\bibinfo {title} {Quantum {Speed} {Limits} across the
  {Quantum}-to-{Classical} {Transition}},}\ }\href
  {https://doi.org/10.1103/PhysRevLett.120.070401} {\bibfield  {journal}
  {\bibinfo  {journal} {Physical Review Letters}\ }\textbf {\bibinfo {volume}
  {120}},\ \bibinfo {pages} {070401} (\bibinfo {year} {2018})}\BibitemShut
  {NoStop}%
\bibitem [{\citenamefont {Funo}, \citenamefont {Shiraishi},\ and\ \citenamefont
  {Saito}(2019)}]{funo_speed_2019}%
  \BibitemOpen
  \bibfield  {author} {\bibinfo {author} {\bibfnamefont {K.}~\bibnamefont
  {Funo}}, \bibinfo {author} {\bibfnamefont {N.}~\bibnamefont {Shiraishi}},\
  and\ \bibinfo {author} {\bibfnamefont {K.}~\bibnamefont {Saito}},\ }\bibfield
   {title} {\enquote {\bibinfo {title} {Speed limit for open quantum
  systems},}\ }\href {https://doi.org/10.1088/1367-2630/aaf9f5} {\bibfield
  {journal} {\bibinfo  {journal} {New Journal of Physics}\ }\textbf {\bibinfo
  {volume} {21}},\ \bibinfo {pages} {013006} (\bibinfo {year}
  {2019})}\BibitemShut {NoStop}%
\bibitem [{\citenamefont {Shiraishi}\ and\ \citenamefont
  {Saito}(2021)}]{shiraishi_speed_2020}%
  \BibitemOpen
  \bibfield  {author} {\bibinfo {author} {\bibfnamefont {N.}~\bibnamefont
  {Shiraishi}}\ and\ \bibinfo {author} {\bibfnamefont {K.}~\bibnamefont
  {Saito}},\ }\bibfield  {title} {\enquote {\bibinfo {title} {Speed limit for
  open systems coupled to general environments},}\ }\href
  {https://doi.org/10.1103/PhysRevResearch.3.023074} {\bibfield  {journal}
  {\bibinfo  {journal} {Phys. Rev. Research}\ }\textbf {\bibinfo {volume}
  {3}},\ \bibinfo {pages} {023074} (\bibinfo {year} {2021})}\BibitemShut
  {NoStop}%
\bibitem [{\citenamefont {Plata}\ \emph {et~al.}(2020)\citenamefont {Plata},
  \citenamefont {Guéry-Odelin}, \citenamefont {Trizac},\ and\ \citenamefont
  {Prados}}]{plata_finite-time_2020}%
  \BibitemOpen
  \bibfield  {author} {\bibinfo {author} {\bibfnamefont {C.~A.}\ \bibnamefont
  {Plata}}, \bibinfo {author} {\bibfnamefont {D.}~\bibnamefont
  {Guéry-Odelin}}, \bibinfo {author} {\bibfnamefont {E.}~\bibnamefont
  {Trizac}},\ and\ \bibinfo {author} {\bibfnamefont {A.}~\bibnamefont
  {Prados}},\ }\bibfield  {title} {\enquote {\bibinfo {title} {Finite-time
  adiabatic processes: {Derivation} and speed limit},}\ }\href
  {https://doi.org/10.1103/PhysRevE.101.032129} {\bibfield  {journal} {\bibinfo
   {journal} {Physical Review E}\ }\textbf {\bibinfo {volume} {101}},\ \bibinfo
  {pages} {032129} (\bibinfo {year} {2020})}\BibitemShut {NoStop}%
\bibitem [{\citenamefont {Deffner}(2020)}]{deffner_quantum_2020}%
  \BibitemOpen
  \bibfield  {author} {\bibinfo {author} {\bibfnamefont {S.}~\bibnamefont
  {Deffner}},\ }\bibfield  {title} {\enquote {\bibinfo {title} {Quantum speed
  limits and the maximal rate of information production},}\ }\href
  {https://doi.org/10.1103/PhysRevResearch.2.013161} {\bibfield  {journal}
  {\bibinfo  {journal} {Physical Review Research}\ }\textbf {\bibinfo {volume}
  {2}},\ \bibinfo {pages} {013161} (\bibinfo {year} {2020})}\BibitemShut
  {NoStop}%
\bibitem [{\citenamefont {Ito}\ and\ \citenamefont
  {Dechant}(2020)}]{ito_stochastic_2020}%
  \BibitemOpen
  \bibfield  {author} {\bibinfo {author} {\bibfnamefont {S.}~\bibnamefont
  {Ito}}\ and\ \bibinfo {author} {\bibfnamefont {A.}~\bibnamefont {Dechant}},\
  }\bibfield  {title} {\enquote {\bibinfo {title} {Stochastic time-evolution,
  information geometry and the {Cramer}-{Rao} {Bound}},}\ }\href
  {https://doi.org/10.1103/PhysRevX.10.021056} {\bibfield  {journal} {\bibinfo
  {journal} {Physical Review X}\ }\textbf {\bibinfo {volume} {10}},\ \bibinfo
  {pages} {021056} (\bibinfo {year} {2020})}\BibitemShut {NoStop}%
\bibitem [{\citenamefont {Van~Vu}\ and\ \citenamefont
  {Hasegawa}(2021)}]{van_vu_geometrical_2021}%
  \BibitemOpen
  \bibfield  {author} {\bibinfo {author} {\bibfnamefont {T.}~\bibnamefont
  {Van~Vu}}\ and\ \bibinfo {author} {\bibfnamefont {Y.}~\bibnamefont
  {Hasegawa}},\ }\bibfield  {title} {\enquote {\bibinfo {title} {Geometrical
  {Bounds} of the {Irreversibility} in {Markovian} {Systems}},}\ }\href
  {https://doi.org/10.1103/PhysRevLett.126.010601} {\bibfield  {journal}
  {\bibinfo  {journal} {Physical Review Letters}\ }\textbf {\bibinfo {volume}
  {126}},\ \bibinfo {pages} {010601} (\bibinfo {year} {2021})}\BibitemShut
  {NoStop}%
\bibitem [{\citenamefont {Prados}(2021)}]{prados_optimizing_2021}%
  \BibitemOpen
  \bibfield  {author} {\bibinfo {author} {\bibfnamefont {A.}~\bibnamefont
  {Prados}},\ }\bibfield  {title} {\enquote {\bibinfo {title} {Optimizing the
  relaxation route with optimal control},}\ }\href
  {https://doi.org/10.1103/PhysRevResearch.3.023128} {\bibfield  {journal}
  {\bibinfo  {journal} {Physical Review Research}\ }\textbf {\bibinfo {volume}
  {3}},\ \bibinfo {pages} {023128} (\bibinfo {year} {2021})}\BibitemShut
  {NoStop}%
\bibitem [{\citenamefont {Lee}\ \emph {et~al.}(2022)\citenamefont {Lee},
  \citenamefont {Lee}, \citenamefont {Kwon},\ and\ \citenamefont
  {Park}}]{lee_speed_2022}%
  \BibitemOpen
  \bibfield  {author} {\bibinfo {author} {\bibfnamefont {J.~S.}\ \bibnamefont
  {Lee}}, \bibinfo {author} {\bibfnamefont {S.}~\bibnamefont {Lee}}, \bibinfo
  {author} {\bibfnamefont {H.}~\bibnamefont {Kwon}},\ and\ \bibinfo {author}
  {\bibfnamefont {H.}~\bibnamefont {Park}},\ }\bibfield  {title} {\enquote
  {\bibinfo {title} {Speed {Limit} for a {Highly} {Irreversible} {Process} and
  {Tight} {Finite}-{Time} {Landauer}’s {Bound}},}\ }\href
  {https://doi.org/10.1103/PhysRevLett.129.120603} {\bibfield  {journal}
  {\bibinfo  {journal} {Physical Review Letters}\ }\textbf {\bibinfo {volume}
  {129}},\ \bibinfo {pages} {120603} (\bibinfo {year} {2022})}\BibitemShut
  {NoStop}%
\bibitem [{\citenamefont {Patrón}, \citenamefont {Prados},\ and\ \citenamefont
  {Plata}(2022)}]{patron_thermal_2022}%
  \BibitemOpen
  \bibfield  {author} {\bibinfo {author} {\bibfnamefont {A.}~\bibnamefont
  {Patrón}}, \bibinfo {author} {\bibfnamefont {A.}~\bibnamefont {Prados}},\
  and\ \bibinfo {author} {\bibfnamefont {C.~A.}\ \bibnamefont {Plata}},\
  }\bibfield  {title} {\enquote {\bibinfo {title} {Thermal brachistochrone for
  harmonically confined {Brownian} particles},}\ }\href
  {https://doi.org/10.1140/epjp/s13360-022-03150-3} {\bibfield  {journal}
  {\bibinfo  {journal} {The European Physical Journal Plus}\ }\textbf {\bibinfo
  {volume} {137}},\ \bibinfo {pages} {1011} (\bibinfo {year}
  {2022})}\BibitemShut {NoStop}%
\bibitem [{\citenamefont {Dechant}(2022)}]{dechant_minimum_2022}%
  \BibitemOpen
  \bibfield  {author} {\bibinfo {author} {\bibfnamefont {A.}~\bibnamefont
  {Dechant}},\ }\bibfield  {title} {\enquote {\bibinfo {title} {Minimum entropy
  production, detailed balance and {Wasserstein} distance for continuous-time
  {Markov} processes},}\ }\href {https://doi.org/10.1088/1751-8121/ac4ac0}
  {\bibfield  {journal} {\bibinfo  {journal} {Journal of Physics A:
  Mathematical and Theoretical}\ }\textbf {\bibinfo {volume} {55}},\ \bibinfo
  {pages} {094001} (\bibinfo {year} {2022})}\BibitemShut {NoStop}%
\bibitem [{Note2()}]{Note2}%
  \BibitemOpen
  \bibinfo {note} {These finite jump discontinuities do not contribute to the
  dissipated heat but make the source voltage have two delta peaks at the
  initial and final times, diverging to $+\infty $ and $-\infty $,
  respectively. This behavior is analogous to that found in the minimization of
  the irreversible work of an underdamped particle under harmonic confinement
  in a thermal bath at temperature $T$.\cite
  {gomez-marin_optimal_2008}}\BibitemShut {NoStop}%
\bibitem [{\citenamefont {Hou}\ \emph {et~al.}(2019)\citenamefont {Hou},
  \citenamefont {Yang}, \citenamefont {He},\ and\ \citenamefont
  {Gao}}]{hou_adaptive_2019}%
  \BibitemOpen
  \bibfield  {author} {\bibinfo {author} {\bibfnamefont {J.}~\bibnamefont
  {Hou}}, \bibinfo {author} {\bibfnamefont {Y.}~\bibnamefont {Yang}}, \bibinfo
  {author} {\bibfnamefont {H.}~\bibnamefont {He}},\ and\ \bibinfo {author}
  {\bibfnamefont {T.}~\bibnamefont {Gao}},\ }\bibfield  {title} {\enquote
  {\bibinfo {title} {Adaptive {Dual} {Extended} {Kalman} {Filter} {Based} on
  {Variational} {Bayesian} {Approximation} for {Joint} {Estimation} of
  {Lithium}-{Ion} {Battery} {State} of {Charge} and {Model} {Parameters}},}\
  }\href {https://doi.org/10.3390/app9091726} {\bibfield  {journal} {\bibinfo
  {journal} {Applied Sciences}\ }\textbf {\bibinfo {volume} {9}},\ \bibinfo
  {pages} {1726} (\bibinfo {year} {2019})}\BibitemShut {NoStop}%
\bibitem [{\citenamefont {Dreyer}, \citenamefont {Guhlke},\ and\ \citenamefont
  {Herrmann}(2011)}]{dreyer_hysteresis_2011}%
  \BibitemOpen
  \bibfield  {author} {\bibinfo {author} {\bibfnamefont {W.}~\bibnamefont
  {Dreyer}}, \bibinfo {author} {\bibfnamefont {C.}~\bibnamefont {Guhlke}},\
  and\ \bibinfo {author} {\bibfnamefont {M.}~\bibnamefont {Herrmann}},\
  }\bibfield  {title} {\enquote {\bibinfo {title} {Hysteresis and phase
  transition in many-particle storage systems},}\ }\href@noop {} {\bibfield
  {journal} {\bibinfo  {journal} {Continuum Mechanics and Thermodynamics}\
  }\textbf {\bibinfo {volume} {23}},\ \bibinfo {pages} {211--231} (\bibinfo
  {year} {2011})}\BibitemShut {NoStop}%
\bibitem [{\citenamefont {Dreyer}, \citenamefont {Guhlke},\ and\ \citenamefont
  {Huth}(2011)}]{dreyer_behavior_2011}%
  \BibitemOpen
  \bibfield  {author} {\bibinfo {author} {\bibfnamefont {W.}~\bibnamefont
  {Dreyer}}, \bibinfo {author} {\bibfnamefont {C.}~\bibnamefont {Guhlke}},\
  and\ \bibinfo {author} {\bibfnamefont {R.}~\bibnamefont {Huth}},\ }\bibfield
  {title} {\enquote {\bibinfo {title} {The behavior of a many-particle
  electrode in a lithium-ion battery},}\ }\href@noop {} {\bibfield  {journal}
  {\bibinfo  {journal} {Physica D: Nonlinear Phenomena}\ }\textbf {\bibinfo
  {volume} {240}},\ \bibinfo {pages} {1008--1019} (\bibinfo {year}
  {2011})}\BibitemShut {NoStop}%
\bibitem [{\citenamefont {Dreyer}\ \emph {et~al.}(2010)\citenamefont {Dreyer},
  \citenamefont {Jamnik}, \citenamefont {Guhlke}, \citenamefont {Huth},
  \citenamefont {Mo\u{s}kon},\ and\ \citenamefont
  {Gaber\u{s}\u{c}ek}}]{dreyer_thermodynamic_2010}%
  \BibitemOpen
  \bibfield  {author} {\bibinfo {author} {\bibfnamefont {W.}~\bibnamefont
  {Dreyer}}, \bibinfo {author} {\bibfnamefont {J.}~\bibnamefont {Jamnik}},
  \bibinfo {author} {\bibfnamefont {C.}~\bibnamefont {Guhlke}}, \bibinfo
  {author} {\bibfnamefont {R.}~\bibnamefont {Huth}}, \bibinfo {author}
  {\bibfnamefont {J.}~\bibnamefont {Mo\u{s}kon}},\ and\ \bibinfo {author}
  {\bibfnamefont {M.}~\bibnamefont {Gaber\u{s}\u{c}ek}},\ }\bibfield  {title}
  {\enquote {\bibinfo {title} {The thermodynamic origin of hysteresis in
  insertion batteries},}\ }\href@noop {} {\bibfield  {journal} {\bibinfo
  {journal} {Nature materials}\ }\textbf {\bibinfo {volume} {9}},\ \bibinfo
  {pages} {448--453} (\bibinfo {year} {2010})}\BibitemShut {NoStop}%
\bibitem [{\citenamefont {Li}\ \emph {et~al.}(2018)\citenamefont {Li},
  \citenamefont {Lu}, \citenamefont {Chen},\ and\ \citenamefont
  {Amine}}]{li_30_2018}%
  \BibitemOpen
  \bibfield  {author} {\bibinfo {author} {\bibfnamefont {M.}~\bibnamefont
  {Li}}, \bibinfo {author} {\bibfnamefont {J.}~\bibnamefont {Lu}}, \bibinfo
  {author} {\bibfnamefont {Z.}~\bibnamefont {Chen}},\ and\ \bibinfo {author}
  {\bibfnamefont {K.}~\bibnamefont {Amine}},\ }\bibfield  {title} {\enquote
  {\bibinfo {title} {30 {Years} of {Lithium}‐{Ion} {Batteries}},}\ }\href
  {https://doi.org/10.1002/adma.201800561} {\bibfield  {journal} {\bibinfo
  {journal} {Advanced Materials}\ }\textbf {\bibinfo {volume} {30}},\ \bibinfo
  {pages} {1800561} (\bibinfo {year} {2018})}\BibitemShut {NoStop}%
\bibitem [{\citenamefont {Wang}\ \emph {et~al.}(2020)\citenamefont {Wang},
  \citenamefont {Tian}, \citenamefont {Sun}, \citenamefont {Wang},
  \citenamefont {Xu}, \citenamefont {Li},\ and\ \citenamefont
  {Chen}}]{wang_comprehensive_2020}%
  \BibitemOpen
  \bibfield  {author} {\bibinfo {author} {\bibfnamefont {Y.}~\bibnamefont
  {Wang}}, \bibinfo {author} {\bibfnamefont {J.}~\bibnamefont {Tian}}, \bibinfo
  {author} {\bibfnamefont {Z.}~\bibnamefont {Sun}}, \bibinfo {author}
  {\bibfnamefont {L.}~\bibnamefont {Wang}}, \bibinfo {author} {\bibfnamefont
  {R.}~\bibnamefont {Xu}}, \bibinfo {author} {\bibfnamefont {M.}~\bibnamefont
  {Li}},\ and\ \bibinfo {author} {\bibfnamefont {Z.}~\bibnamefont {Chen}},\
  }\bibfield  {title} {\enquote {\bibinfo {title} {A comprehensive review of
  battery modeling and state estimation approaches for advanced battery
  management systems},}\ }\href {https://doi.org/10.1016/j.rser.2020.110015}
  {\bibfield  {journal} {\bibinfo  {journal} {Renewable and Sustainable Energy
  Reviews}\ }\textbf {\bibinfo {volume} {131}},\ \bibinfo {pages} {110015}
  (\bibinfo {year} {2020})}\BibitemShut {NoStop}%
\bibitem [{\citenamefont {Manthiram}(2020)}]{manthiram_reflection_2020}%
  \BibitemOpen
  \bibfield  {author} {\bibinfo {author} {\bibfnamefont {A.}~\bibnamefont
  {Manthiram}},\ }\bibfield  {title} {\enquote {\bibinfo {title} {A reflection
  on lithium-ion battery cathode chemistry},}\ }\href
  {https://doi.org/10.1038/s41467-020-15355-0} {\bibfield  {journal} {\bibinfo
  {journal} {Nature Communications}\ }\textbf {\bibinfo {volume} {11}},\
  \bibinfo {pages} {1550} (\bibinfo {year} {2020})}\BibitemShut {NoStop}%
\bibitem [{Note3()}]{Note3}%
  \BibitemOpen
  \bibinfo {note} {Note that the notation employed here is slightly different
  from that of the main text: here, $\varepsilon _{i}$ and $\varepsilon _{f}$
  always stand for the initial and final voltage of the corresponding process,
  whereas in the main text we only introduced $\epsilon _{f}$ as the final
  voltage of the charging process, which was the initial one for the subsequent
  discharging process.}\BibitemShut {Stop}%
\bibitem [{\citenamefont {Gomez-Marin}, \citenamefont {Schmiedl},\ and\
  \citenamefont {Seifert}(2008)}]{gomez-marin_optimal_2008}%
  \BibitemOpen
  \bibfield  {author} {\bibinfo {author} {\bibfnamefont {A.}~\bibnamefont
  {Gomez-Marin}}, \bibinfo {author} {\bibfnamefont {T.}~\bibnamefont
  {Schmiedl}},\ and\ \bibinfo {author} {\bibfnamefont {U.}~\bibnamefont
  {Seifert}},\ }\bibfield  {title} {\enquote {\bibinfo {title} {Optimal
  protocols for minimal work processes in underdamped stochastic
  thermodynamics},}\ }\href {https://doi.org/10.1063/1.2948948} {\bibfield
  {journal} {\bibinfo  {journal} {The Journal of Chemical Physics}\ }\textbf
  {\bibinfo {volume} {129}},\ \bibinfo {pages} {024114} (\bibinfo {year}
  {2008})}\BibitemShut {NoStop}%
\end{thebibliography}%

\end{document}